\documentclass[12pt]{article}
\usepackage{graphicx}

\input{epsf}

\newcommand{\nc}{\newcommand}
\nc{\beq}{\begin{equation}}
\nc{\eeq}{\end{equation}}
\nc{\beqa}{\begin{eqnarray}}
\nc{\eeqa}{\end{eqnarray}}

\def\gsim{\mathrel{\rlap{\lower4pt\hbox{\hskip1pt$\sim$}}
    \raise1pt\hbox{$>$}}}       

\begin{document}



\title{\large{\bf Space-Time Symmetries of Noncommutative Spaces}}

\author{Xavier~Calmet\thanks{calmet@physics.unc.edu} \\
Department of Physics and Astronomy \\
University of North Carolina at Chapel Hill \\
Chapel Hill, NC 27599}

\date{March 27, 2005}

\maketitle

\begin{abstract}
We define a noncommutative Lorentz symmetry for canonical
noncommutative spaces.  The noncommutative vector fields and the
derivatives transform under a deformed Lorentz transformation. We show that the star product is invariant under noncommutative Lorentz transformations. We then
apply our idea to the case of actions obtained by expanding the star
product and the fields taken in the enveloping algebra via
the Seiberg-Witten maps and verify that these actions are invariant
under these new noncommutative Lorentz transformations. We finally
consider general coordinate transformations and show that the metric
is undeformed.
\end{abstract}

to appear in Phys. Rev. D.


\newpage


Lorentz symmetry plays a central role in any realistic quantum field
theory. Recently, due to progress in string/M theory
\cite{Connes:1997cr}, the idea that space-time could involve at short
distances some non trivial noncommutative coordinates was revived. But
these quantum field theories typically violate Lorentz
invariance. In the first paper on space-time noncommutativity
\cite{Snyder:1946qz}, Snyder argued that Lorentz invariance is not
incompatible with a discrete space-time and he gave a concrete
noncommutative algebra that allows to recover Lorentz invariance, but
not Poincar\'e invariance.

 Noncommutative gauge theories are very interesting since
they represent simple examples of models with a minimal length and it
has recently been established that quantum mechanics considered
together with classical general relativity imply the existence of a
minimal length in nature \cite{Calmet:2004mp}. Nevertheless gauge
theories formulated on a canonical noncommutative space-time violate
Lorentz invariance. Although it is known how to formulate the standard
model on a noncommutative space-time \cite{Calmet:2001na} (see also
\cite{Chaichian:2004yw} for another approach), there is no obvious way
to preserve Lorentz invariance and the bounds on the noncommutative
nature of space-time are actually derived from bounds on Lorentz
invariance violation \cite{Calmet:2004dn}. One way to consider Lorentz
invariant noncommutative models is to consider space-time dependent
noncommutativity \cite{Calmet:2003jv}, but this approach has not yet
been studied in great details and it remains a speculation. In this
work we show that noncommutative theories formulated on a canonical
space-time have an underlying exact symmetry that corresponds to
Lorentz invariance in the limit $\theta^{\mu\nu} \to 0$. We call this
symmetry noncommutative Lorentz invariance.  Let us consider the
noncommutative algebra:
\begin{eqnarray} \label{NCA}
[ \hat x^i,\hat x^j]=i\theta^{ij}
\end{eqnarray}
where $i,j$ run from 1 to 3 and where we set $\theta^{0i}=0$, i.e. we
assume that the space coordinates commute with the time coordinate. It
will soon become obvious why we restrict our considerations to that
case. Furthermore one has the Heisenberg algebra:
\begin{eqnarray}
[ \hat x^i,p^j]=i \hbar \delta^{ij}
\end{eqnarray}
and
\begin{eqnarray}
[  p^i,p^j]=0.
\end{eqnarray}
One could try to introduce a noncommutative Lorentz symmetry by
imposing a transformation $\hat x^i= \Lambda^i_{\ j} \hat x^j$, but
that is not consistent with the algebra (\ref{NCA}) since it would
require that $\theta^{ij}$ transforms \footnote{ This approach has been considered in \cite{Doplicher:1994tu}, but we wish to treat $\theta^{ij}$ as a universal constant tensor, just like the speed of light in special relativity on commutative spaces.} as $\Lambda^i_{\ k} \Lambda^i_{\
l} \theta^{kl}$ which makes little sense since it is by definition a
constant and thus should remain invariant.

It is easy to see that one can introduce a new operator $x_c^i$ defined by
\begin{eqnarray} \label{CLT}
x_c^i=\hat x^i +\frac{1}{2\hbar} \theta^{ij} p_j,
\end{eqnarray}
 which leads to the following algebras:
\begin{eqnarray} \label{NAlg}
[ x_c^i, x_c^j]=0, \ [  x_c^i,p^j]=i \hbar \delta^{ij} \ \mbox{and} \ [  p^i,p^j]=0,
\end{eqnarray}
i.e. $x_c^i$ are commuting coordinates.  Since $t$ is not an operator
in quantum mechanics, one cannot eliminate the constraint (\ref{NCA})
for space-time noncommutativity, this explains our previous assumption
$\theta^{0i}=0$.  This condition has to be imposed in the string/M
theory approach \cite{Douglas:2001ba} to noncommutative gauge theories
to avoid problems with unitarity \cite{Gomis:2000zz}. We can now treat
the problem in a covariant way and introduce Greek variables which are
running from 0 to 3. Given the algebras (\ref{NAlg}), we can define a transformation
\begin{eqnarray} 
x_c^\mu= \Lambda^\mu_{\ \nu} x_c^\nu
\end{eqnarray}
that leaves the interval $s^2= \eta_{\mu\nu} x_c^\mu x_c^\nu$
invariant if $\eta_{\mu\nu} \Lambda^\mu_{\ \alpha} \Lambda^\nu_{\
\beta} = \eta_{\alpha \beta} $. Notice that $p^\mu$ transforms as an
usual Lorentz vector, i.e.
\begin{eqnarray} 
p^\mu= \Lambda^\mu_{\ \nu} p^\nu.
\end{eqnarray}
One thus finds that the transformation (\ref{CLT}), implies that $\hat x^\mu$ transforms as
\begin{eqnarray}  
\hat x^{\mu \prime} &=& x_c^{\mu \prime} - \frac{1}{2\hbar}
\theta^{\mu\nu} p_\nu^\prime= \Lambda^\mu_{\ \nu} x_c^\nu -
\frac{1}{2\hbar} \theta^{\nu \rho} \Lambda_\rho^{\ \sigma} p_{\sigma}
\end{eqnarray}
or
\begin{eqnarray} \label{NLT}
\hat x^{\mu\prime} &=& \Lambda^\mu_{\ \nu} \hat x^\nu + \frac{1}{2\hbar}   \Lambda^\mu_{\ \nu} \theta^{\nu \rho} p_\rho 
- \frac{1}{2\hbar}  \theta^{\mu\nu} \Lambda_\nu^{\ \rho} p_{\rho} 
\end{eqnarray}
which defines the noncommutative Lorentz transformation. Note that the
second term $\Lambda^\mu_{\ \nu} \theta^{\nu\rho} p_\rho$ is not a
transformation of the noncommutative parameters $\theta^{\mu\nu}$, but
of $\theta^{\mu\nu}p_\nu$. It is easy to verify that the algebra
(\ref{NCA}) is left invariant by this transformation and that there is
a smooth limit to the Lorentz transformation on classical space-time
by taking the limit $\theta^{\mu\nu} \to 0$. We now define the
noncommutative invariant length. The square of the invariant length
for the commutative coordinate $x^\mu_c$ is
\begin{eqnarray} 
s^2= \eta_{\mu\nu} x^\mu_c x^\nu_c.
\end{eqnarray}
Using the variable transformation (\ref{CLT}), one finds that the square of the noncommutative invariant length is given by
\begin{eqnarray} 
s^2_{nc}= \hat x^\mu \hat x_\mu + \frac{1}{\hbar} \theta_{\mu\nu} \hat
x^\mu p^\nu + \frac{1}{4\hbar^2} \theta^{\mu\alpha} \theta_{\mu \beta}
p_\alpha p^\beta.
\end{eqnarray}
It is easy to verify that $s^2_{nc}$ is left invariant by the noncommutative Lorentz transformation 
(\ref{NLT}). This is the way we define the noncommutative Lorentz transformations, those are the transformations that leave $s^2_{nc}$ invariant.

It is straightforward to extend our results to a Poincar\'e
transformation since a shift by a constant of the noncommutative
coordinates is compatible with the algebra (\ref{NCA}). Let us now
consider an infinitesimal noncommutative Poincar\'e transformation
$\Lambda^\mu_{\ \nu}=\delta^\mu_{\ \nu}+\omega^\mu_{\ \nu},
a^\mu=\epsilon^\mu$. It is implemented by the operator:
\begin{eqnarray}
U(1+\omega,\epsilon)=1 +\frac{1}{2} i \omega_{\rho\sigma} J^{\rho\sigma} - i
\epsilon_\rho p^\rho + ...
\end{eqnarray}
with $J^{\mu\nu}=x_c^\mu p^\nu -x_c^\nu p^\mu$. The operator is
undeformed. The Lie algebra  of the Lorentz group is  also
undeformed:
\begin{eqnarray}
i [J^{\mu\nu}, J^{\rho\sigma}]&=& \eta^{\nu\rho} J ^{\mu\sigma} - \eta^{\mu\rho} J ^{\nu\sigma}
-\eta^{\sigma\mu} J ^{\rho\nu} +\eta^{\sigma\nu} J ^{\rho\mu}   \\
i[p^\mu,J^{\rho\sigma}]&=&\eta^{\mu\rho}p^\sigma-\eta^{\mu\sigma}p^\rho  \\
\ [ p^\mu , p^\rho ] &=&0. 
\end{eqnarray}
We note that our approach is different from the twisted Poincar\'e
symmetry considered in \cite{Wess:2004da}. It is also different from
the $\kappa$-Poincar\'e quasi group where the Poincar\'e symmetry is
deformed \cite{Lukierski:1992dt}.

We shall now consider field theories. We need to introduce a
derivative. Derivatives have to be defined in such a way, that they do
not lead to new relations for the coordinates. In the canonical case,
it is easy to show that $\hat x^\alpha - i
\theta^{\alpha\rho}\hat\partial_\rho$ with $\hat \partial_\rho \hat
x^\mu=\delta^\mu_\rho +\hat x^\mu \hat \partial_\rho$ commutes with
all coordinates \cite{Jurco:2000ja}. One thus finds $\hat \partial_\mu f=
-i \theta^{-1}_{\mu \nu} [\hat x^\nu,f]$. In our case we need a derivative
which is compatible with the noncommutative Lorentz symmetry. We
define the derivative in the following way:
\begin{eqnarray}
i \theta_{\mu\nu} \hat \partial^\nu f = 2 [ \hat x_\mu +\frac{1}{2 \hbar} \theta_{\mu\alpha} p^\alpha, f ]
\end{eqnarray}
with $[p^\mu,f]=-i\hbar \partial^\mu f$. Note that the left hand side
of the equation is covariant. One finds that the derivative
$\hat \partial_\nu$ transforms as
\begin{eqnarray}
\hat \partial_\nu^\prime= \theta^{-1}_{\nu \alpha} \Lambda^{\alpha}_{\ \beta} \theta^{\beta \rho} \hat \partial_\rho
\end{eqnarray}
under a noncommutative Lorentz transformation. 
We can thus write a noncommutative Lorentz invariant free field action for a noncommutative scalar field:
 \begin{eqnarray}
S=\int d^4x \left (\hat \partial_\mu \Phi \hat \partial^\mu \Phi -m^2 \Phi \Phi - \lambda \Phi \Phi \Phi \Phi \right). 
\end{eqnarray}
Note that the one-particle states are classified according to the eigenvectors of the four-momentum which transforms as usual under Lorentz transformations. The scalar, vector and spinor fields thus transform in the usual way under Lorentz transformations.
The Weyl quantization procedure can be applied to map the noncommutative fields $\Phi(\hat x)$ to the commutative ones $\Phi(x)$. As usual, this corresponds to a replacement of the multiplication operation by a star product given by $f(x) \star g(x) = f(x) \exp(-i \partial_\mu \theta^{\mu\nu} \partial_\nu) g(x)$. It is easy to verify that the star product  is invariant under noncommutative Lorentz transformations. The noncommutative gauge theories inspired by string theory are thus invariant under these transformations.

The noncommutative Lorentz transformation is compatible with gauge
transformations. Remember that one has to introduce a covariant
coordinate $\hat X^\mu$ \cite{Madore:2000en} such that $\hat \delta_{\hat
\Lambda} (\hat X^\mu \hat \Psi(\hat x))=\hat \Lambda \hat X^\mu \hat \Psi(\hat
x)$ where $\hat \Lambda$ is a noncommutative gauge transformation. One
finds that $\hat X^\mu=\hat x^\mu+\hat B^\mu$ with $\hat \delta_{\hat
\Lambda} \hat B^\mu=i [\hat \Lambda, \hat B^\mu] -i[\hat x^\mu, \hat
\Lambda]$. The Yang-Mills gauge potential $\hat A^\mu$ is related to
the gauge potential for the coordinate $\hat B^\mu$ by the relation
$\hat B^\mu=\theta^{\mu \nu} \hat A_\nu$ and the covariant derivative
$\hat D^\mu$ is given by $\hat D_\mu=-i \theta^{-1}_{\mu\nu} \hat
X^\nu $. The coordinate gauge potential $\hat B_\mu$ transforms
as $\hat B^\prime_\mu = \Lambda_\mu^{\ \nu} \hat B_\nu$, one thus finds
that the noncommutative Yang-Mills potential transforms as 
 \begin{eqnarray}
\hat A^\prime_\mu = \theta^{-1}_{\mu\nu} \Lambda^\nu_{\ \rho} \theta^{\rho
\sigma} \hat A_\sigma. 
\end{eqnarray}
The noncommutative covariant derivative transforms as
\begin{eqnarray} \label{T1}
\hat D_{\mu}^\prime= \theta^{-1}_{\mu \rho} \Lambda^\rho_{\ \sigma}
\theta^{\sigma \alpha} \hat D_{\alpha}
\end{eqnarray}
under a noncommutative Lorentz transformation.
 The field strength $\hat F_{\mu\nu}$ is given by  $\hat F_{\mu\nu}= i [\hat D_\mu,\hat D_\nu]$, it  transforms as 
\begin{eqnarray} \label{T2}
\hat F_{\mu\nu}^\prime= \theta^{-1}_{\mu \rho} 
\Lambda^\rho_{\ \sigma} \theta^{\sigma \alpha}
                              \theta^{-1}_{\nu \kappa} \Lambda^\kappa_{\ \xi} \theta^{\xi \beta}  \hat F_{\alpha\beta}
\end{eqnarray}
under a noncommutative Lorentz transformation.
The noncommutative spinor field $\hat \Psi$ transforms as 
\begin{eqnarray} \label{T3}
\hat \Psi^\prime= \exp \left ( -\frac{i}{2} w^{\alpha \beta} S_{\alpha \beta} \right  )  \hat \Psi,
\end{eqnarray}
with $S^{\mu\nu}=\frac{i}{4}[\gamma^\mu,\gamma^\nu]$.  Note that if
the fields are taken in the enveloping algebra, the leading
order field of the Seiberg-Witten expansion \cite{Seiberg:1999vs},
i.e. the classical field, also transforms according to (\ref{T1}),
(\ref{T2}) and (\ref{T3}).

Up to this point our considerations were completely general and did not
assume a specific approach to space-time noncommutativity. We now
apply our results to a specific framework, namely we consider fields
to be in the enveloping algebra. Given (\ref{T1}),
(\ref{T2}) and (\ref{T3}) is it easy to verify that the effective
action obtained in the leading order in $\theta$ after the expansion
of the noncommutative fields via the Seiberg-Witten map and of the
star product:
\begin{eqnarray} \label{action}
S &=& \int d^4 x \bigg[ \bar \psi (i \! \not \! \! {D} - m
)\psi - \frac 1 4 \, \theta^{\mu \nu} \bar \psi F_{\mu \nu}
(i \not \! \! {D} - m )\psi - \frac{1}{2} \, \theta^{\mu \nu}
\bar \psi \gamma^\rho F_{\rho \mu} \, i {D}_\nu \psi
\\ \nonumber &&
 -\frac{1}{2} \, {\rm Tr} \, F_{\mu \nu} F^{\mu \nu}
+ \frac{1}{4}  \, \theta^{\mu \nu} \, {\rm Tr} \, F_{\mu \nu}
F_{\rho \sigma} F^{ \rho \sigma} -  \, \theta^{\mu
\nu} \, {\rm Tr} \, F_{\mu \rho} F_{\nu \sigma} F^{ \rho
\sigma} \bigg] + {\cal O}(\theta^2)
\end{eqnarray}
is invariant under noncommutative Lorentz transformations.  

There are implications for the bounds on space-time noncommutativity
\cite{Carroll:2001ws,Calmet:2004dn}. The 10 TeV bound on space-time
noncommutativity when fields are taken in the enveloping algebra comes from atomic clock comparison studies. These studies
search for a difference between two atomic transition frequencies,
searching for variations as the Earth rotates \cite{Prestage:1985zm}.
The 10 TeV bound was obtained in \cite{Carroll:2001ws} assuming that
the fermionic sector of (\ref{action}) transforms according to the
classical Lorentz transformations.  If we posit that the noncommutative Lorentz invariance is a symmetry of nature, one should  use the noncommutative Lorentz transformations described in this work  to compare the laboratory frame to the laboratory frame rotating with the Earth. The  bounds on space-time noncommutativity coming from atomic clock comparison have to be reconsidered.  A noncommutative Lorentz transformation corresponding to a $2 \pi$ rotation would not take a system back to the same point, one could imagine testing this symmetry by measuring the spectrum of some transition in e.g. a nuclei and by studying how the spectrum is affected if the complete experiment is rotated by $2 \pi$. We emphasize that in our case $\theta$ is a constant in all reference frames, i.e. our symmetry is not spontaneously broken.  Tests of Lorentz invariance, in the framework of noncommutative gauge theories, usually assume that $\theta$ changes  from one reference frame to another and thus breaks Lorentz invariance spontaneously. It is thus not obvious how to use bounds on spontaneous violations of Lorentz to  constrain our symmetry. A detail analysis of the phenomenological consequences of this symmetry will appear elsewhere and is beyond the scope of this article.

The noncommutative Lorentz symmetry  has also implications for the bounds relevant to the string/M theory approach to space-time noncommutativity. In that case the bounds come from the operators $O_1= m_e \theta^{\mu\nu} \bar \psi \sigma_{\mu\nu} \psi$, $O_2=  \theta^{\mu\nu} \bar \psi D_\mu \gamma_{\nu} \psi$,  $O_3= \lambda_3 \theta^{\mu\alpha} \theta_{\alpha\nu} F_{\mu\rho} F^{\rho\nu}$, $O_4=\lambda_4/8 (\theta^{\mu\nu} F_{\mu\nu})^2$ and $O_5=\theta_{\mu\rho} F^{\rho\sigma} \theta_{\sigma\gamma} F^{\gamma \mu}$ which are typically generated at two loops   \cite{Anisimov:2001zc}. It is however easy to verify that these operators are not invariant under the transformations (\ref{T2}) and (\ref{T3}) and are thus an artifact of the cutoff used to regularized the divergent integrals. In that case again, the bounds on space-time noncommutativity are affected if we postulate that the noncommutative Lorentz symmetry is a symmetry of nature. On the other hand one finds that the effective cutoff responsible for the UV/IR phenomenon: $\Lambda^2_{eff}=(1/\Lambda^2-p_\mu \theta^2_{\mu\nu} q_\nu)^{-1}$ \cite{Minwalla:1999px} is invariant under the deformed Lorentz symmetry. The UV/IR mixing phenomenon is thus not related to a symmetry of the noncommutative space-time.

It is straightforward to extend our results to the case of  general coordinate transformations. As for the case of Lorentz transformations, we can consider general coordinate transformations of the commutative variable $x^\mu_c$. One finds that the infinitesimal length interval
\begin{eqnarray} \label{metric}
ds^2=g_{\mu\nu}(x) dx_c^\mu dx^\nu_c
\end{eqnarray}
is invariant under a general coordinate transformation $x_\mu \to \xi_\mu$, if the metric transforms as $g_{\mu\nu}=g_{\alpha\beta} \frac{\partial x^\alpha}{\partial \xi^\mu}\frac{\partial x^\beta}{\partial \xi^\nu}$. Applying the variable transformation  (\ref{CLT}) to the infinitesimal length interval, we find
\begin{eqnarray} 
ds^2=g_{\mu\nu}(x) \frac{\partial x_c^\mu}{\partial \hat x^\alpha}  d\hat x^\alpha \frac{\partial x_c^\nu}{\partial \hat x^\beta} d\hat x^\beta.
\end{eqnarray}
Using $\frac{\partial x_c^\mu}{\partial \hat x^\alpha}=-i 2 \theta^{-1}_{\alpha\nu}  [ \hat x^\nu +\frac{1}{2 \hbar} \theta^{\nu\sigma} p_\sigma, x_c^\mu ]=\delta^\mu_{\ \alpha}$, we find
\begin{eqnarray}
 ds^2=g_{\mu\nu}(\hat x) d\hat x^\mu d\hat x^\nu.
 \end{eqnarray}
The noncommutative metric is therefore undeformed. This does not imply that the noncommutative Einstein action will itself be undeformed \cite{calmet}.

In summary we have defined space-time transformations for noncommutative spaces. The basic idea is to define these transformations for a commutative variable and to feed back these transformations to the noncommutative sector via a variable transformation. We have shown that the $\theta$-expanded action is invariant under noncommutative Lorentz transformations and we have applied the same idea to general coordinate transformations and shown that the metric remains undeformed, this might not be a surprise since in string/M theory, gravity is determined by closed strings that do not feel the noncommutativity. 
\bigskip
\subsection*{Acknowledgments}
\noindent 
The author would like to thank Stephen Hsu for an interesting
discussion and the Institute of Theoretical Science at the University of Oregon where this work
was started for its hospitality. He would also like to thank Archil
Kobakhidze for helpful comments and
suggestions. This work was supported in part by the US Department of
Energy under Grant No. DE-FG02-97ER-41036.


\bigskip


\end{document}